\documentstyle[preprint,aps,epsfig,eqsecnum]{revtex}

\newcommand{\MB}{m_B}
\newcommand{\MD}{m_D}

\newcommand{\be}{\begin{eqnarray}}
\newcommand{\ee}{\end{eqnarray}}
\newcommand{\bl}{$B_{l4}$ }

\renewcommand{\Im}{{\rm Im}}

\tightenlines
\begin{document}

\newpage
\setcounter{page}{0}

\begin{titlepage}
 \begin{flushright}
 \hfill{hep-ph/9811396}\\
 \hfill{YUMS 98--020}\\
 \hfill{SNUTP 98--127}\\
 \hfill{KIAS-P98036}\\

 \end{flushright}
\vspace*{0.8cm}

\begin{center}
{\large\bf CP Violation in the Semileptonic $B_{l4}$ ($B \rightarrow D \pi l \nu$) 
 Decays: \\
 Multi-Higgs Doublet Model and Scalar-Leptoquark Models}
\end{center}
\vskip 0.8cm
\begin{center}
{\sc C. S.~Kim$^{\mathrm{a,b,}}$\footnote{e-mail : kim@cskim.yonsei.ac.kr,
~~ http://phya.yonsei.ac.kr/\~{}cskim/},
 Jake Lee$^{\mathrm{a,}}$\footnote{e-mail : jilee@theory.yonsei.ac.kr} and 
 W.~Namgung$^{\mathrm{c,}}$\footnote{e-mail : ngw@cakra.dongguk.ac.kr}}

\vskip 0.5cm

\begin{small} 
$^{\mathrm{a}}$ Department of Physics, Yonsei University 120-749, Seoul,   
                  Korea. \\
\vskip 0.2cm
$^{\mathrm{b}}$ School of Physics, Korea Institute for Advanced Study, 
            Seoul 130-012, Korea \\
\vskip 0.2cm
$^{\mathrm{c}}$ Department of Physics, Dongguk University 100-715, 
                  Seoul, Korea.
\end{small}
\end{center}

\vspace{0.5cm}
\begin{center}
 (\today)
\end{center}

\setcounter{footnote}{0}
\begin{abstract}
CP violation from physics beyond the Standard Model is investigated 
in $B_{l4}$ decays: $B\to D\pi l\bar{\nu}_l$. 
In the decay process, we include various excited states as intermediate states 
decaying to the final hadrons, $D+\pi$. 
We consider the semileptonic decay to a tau lepton family as well.
The CP violation is implemented through complex scalar--fermion couplings
in the multi-Higgs doublet model and scalar-leptoquark models 
beyond the Standard Model.
With these complex couplings,
we calculate the CP-odd rate asymmetry and the optimal asymmetry. 
We find that for $B_{\tau 4}$ decays the optimal asymmetry is sizable and
can be detected at $1\sigma$ level with about $10^6$-$10^7$ $B$-meson pairs, 
for maximally-allowed values of CP-odd parameters in those extended models.
\end{abstract}
\vskip 1cm
%
\end{titlepage}

\newpage
\baselineskip .29in
\renewcommand{\thefootnote}{\alph{footnote}}

\section{Introduction}

\noindent
The pions emitted in $B_{l4}$ decays ($B\to D\pi l\bar{\nu}_l$) 
have such a wide momentum range that
one may have difficulties to analyze the decays over whole phase space.
However, if we restrict our attentions to the soft pion limit,
we could investigate $B_{l4}$ decays including the final state $D$-meson  
by the combined method of heavy quark expansion and chiral 
perturbation expansion \cite{BL,goity}.
So far, the significance of $B\to D\pi l\nu$ decay mode has been seen 
from the observation that the elastic modes $B\to De\nu$ and 
$B\to D^*e\nu$ account for less than $70\%$ of the total semileptonic branching 
fraction.
Goity and Roberts \cite{goity} have studied $B\to D\pi l\nu$ decays including
various intermediate states which are decaying to $D+\pi$,
and found that the effects of higher excited intermediate
states are substantial compared to the lowest state of $D^*$, 
as the invariant mass of $D+\pi$ grows away from the ground $D^*$ 
resonance regions.

In Ref.~ \cite{bcp}, we considered the possibility of probing  
direct CP violation in the decay of $B\to D\pi l\nu$ 
in a model independent way, 
in which we extended leptonic current
by including complex couplings  of the scalar sector and those of the vector sector
in extensions of the Standard Model (SM).
In the present paper, we would like to consider specific models 
such as mult-Higgs doublet model and scalar-leptoquark models.
In order to observe direct CP violation effects, 
there should exist interferences not only through weak CP-violating
phases but also with different CP-conserving strong phases.
In \bl decays, CP-violating phases can be generated through interference
between $W$-exchange diagrams and 
Scalar-exchange diagrams with complex couplings. 
The CP-conserving phases may come from the absorptive parts of the intermediate
resonances in \bl decays. 

We include higher excited states, such as $P$-wave, $D$-wave states 
and the first radially excited $S$-wave
state, as intermediate states in $B\to D\pi l\nu$ decay, since these
intermediate states could contribute substantially to the decay \bl \cite{goity}.
And we have found in our previous work \cite{bcp} that CP-violation effects are 
highly amplified by including those higher excited states.

The interactions of the octet of pseudogoldstone bosons with hadrons containing 
a single
heavy quark are constrained by two independent symmetries: 
spontaneously broken chiral
SU(3)$_L\times $ SU(3)$_R$ symmetry and heavy quark spin-flavor SU$(2N_h)$ 
symmetry \cite{HQET},
where $N_h$ is the number of heavy quark flavors.
Within the frame work of heavy quark effective theory (HQET) \cite{spect},
the spin $J$ of a meson consists of the spin of a heavy quark ($Q$),
the spin of a light quark ($q$) and relative angular momentum $l$:
\be
J=S_Q+S_l+l\;.
\ee
We denote the meson state as $J^P$ with its parity.
If we define $j=S_l+l$ which corresponds to the spin of the light component of 
the meson, we have a multiplet for each $j$:
\be
J=j\pm \frac{1}{2}\;.
\ee
Then, for a meson $M$, HQET predicts the following multiplets up to $l=2$:
\be
l=0:&& (0^-,1^-)\qquad (M,M^*)\;,\nonumber\\
l=1:&& (0^+,1^+)\qquad (M_0,M_1^{(0)})\nonumber\\
    && (1^+,2^+)\qquad (M_1^{(1)},M_2)\;,\nonumber\\
l=2:&& (1^-,2^-)\qquad (M_1,M_2^{(0)})\nonumber\\
    && (2^-,3^-)\qquad (M_2^{(1)},M_3)\;,
\ee
where we denote the corresponding meson states by the notation in the last column.
Furthermore, there could be radially excited states.
For example, the first radially excited states are
\be
l=0\;(n=2):\; (0^-,1^-)\qquad (M',M'^*)\;.
\ee
Among these resonances (denoted by $\tilde{M}$), $\tilde{D}\to D\pi$ or
$B\to \tilde{B}\pi$ decay is possible only for $J^P=0^+$, $1^-$, $2^+$ and $3^-$
resonances because of parity conservation.
However, if we use chiral expansion, the decay amplitude of $2^+$ state, $M_2$, is 
proportional to $(p_\pi)^2$, and 
that of $3^-$ state is proportional to $(p_\pi)^3$  \cite{goity}, 
so their contributions will be suppressed in the soft pion limit.
Therefore, in the leading order in $p_\pi$, the resonances contributing
to $B_{l4}$ decays are
\be
\tilde{M}=M^*,\; M_0,\; M_1\;{\rm and}\; M'^*,
\ee
where $M$ stands for $B$ or $D$ meson.
We include all the possible multipets above as intermediate states in \bl decays.

In Section II, we briefly review on our formalism dealing with \bl decays.
The multi-Higgs doublet model and scalar-leptoquark models are introduced as
new sources of CP violation in Section III, and the observable asymmetries are
considered in Section IV.
Section V contains our numerical results and conclusions.

\section{Decay Rates}

\noindent
We consider $B_{l4}$ decays of $B^-\to D^+\pi^-l\bar{\nu}$ and
$B^-\to D^0\pi^0l\bar{\nu}$.
The amplitude  has the general form:
\be
T=\kappa \Big[(1+\chi)j_\mu\Omega_V^\mu+\eta j_s \Omega_s\Big],\qquad 
\kappa=V_{cb}\frac{G_F}{\sqrt{2}}\sqrt{\MB\MD},
\ee
where $j_\mu$ is the $(V-A)$ charged leptonic current and $j_s$ is 
a Yukawa type scalar current in leptonic sector.
Here the parameters $\chi$ and $\eta$, which parametrize contributions from physics
beyond the SM are in general complex. Note that the SM values are
$\chi=\eta =0$.
We retain charged lepton mass since we also consider decays of $B_{\tau 4}$.
The vector interaction part of the hadronic amplitude, $\Omega_V^\mu$, 
receives contributions from two types of
diagrams, illustrated in Fig.~1(a) and 1(b), respectively:
\be
\Omega_V^\mu&=&\langle D\pi|\bar{c}\gamma^\mu(1-\gamma_5)b|B\rangle \nonumber\\
    &=&\sum_{\tilde{B}_i}\langle D|\bar{c}\gamma^\mu(1-\gamma_5)b|\tilde{B}_i\rangle 
          \langle \tilde{B}_i\pi|B\rangle 
          +\sum_{\tilde{D}_i}\langle D\pi|\tilde{D}_i\rangle \langle \tilde{D}_i|
          \bar{c}\gamma^\mu(1-\gamma_5)b|B\rangle ,
\ee
and $\Omega_s$ is the corresponding scalar current matrix element:
\be
\Omega_s=\langle D\pi|\bar{c}(1-\gamma_5)b|B\rangle ,
\ee
where $\tilde{D}_i$ and $\tilde{B}_i$ stand for intermediate excited states of
our interest.
One can obtain Yukawa interaction form factors by multiplying the $V-A$
currents with momentum transfer $q^\mu = p_B-p_D=\MB v^\mu - \MD v'^\mu$.
Consequently, we get the following relation
\be
q_\mu \Omega_V^\mu=(m_B+m_D)\Omega_s.
\ee
We define the dimensionless parameter $\zeta$, 
which determines the relative size of the 
scalar contributions to the vector ones:
\be
\zeta=\frac{\eta}{1+\chi}.
\label{zeta}
\ee
Then the amplitude can be written as
\be
T=\kappa(1+\chi)j_\mu\Omega^\mu,\qquad \Omega^\mu=\Omega_V^\mu+\zeta\Omega_s\frac{L^\mu}{m_l},
\ee
where we used the Dirac equation for leptonic current,
$L^\mu j_\mu =m_l j_s$, with $L^\mu=p_l^\mu+p_{\bar{\nu}}^\mu$.
All the explicit expressions of the form factors and the amplitudes can be 
found in Ref.~\cite{bcp}.

Then, the differential partial width of interest can be expressed as
\be
d\Gamma_{B_{l4}}=\frac{N_\pi}{2\MB}J(s_M,s_L)|T|^2 d \Phi_4,
\ee
where the 4 body phase space $d \Phi_4$ is
\be
d \Phi_4 \equiv  ds_M \cdot ds_L \cdot d\cos\theta \cdot d\cos\theta_l \cdot d\phi,
\ee
and
\be
N_\pi = \left\{\begin{array}{cc}
            2 & {\rm for\; charged\; pions,}\\
            1 & {\rm for\; neutral\; pions,}
            \end{array}\right.
\ee
and the Jacobian $J(x,y)$ is
\be
J(x,y)=\frac{1}{2^{14}\pi^6xy\MB^2}\lambda^{1/2}(\MB^2,x,y)
      \lambda^{1/2}(x,\MD^2,m_\pi^2)\lambda^{1/2}(y,m_l^2,0).
\ee
Here, from the momenta of the $B$-meson,
$D$-meson, the pion, the lepton, and its neutrino $p_B$, $p_D$, $p$, $p_l$, 
and $p_{\bar{\nu}}$, for five independent kinematic variables we choose 
$s_M=(p_D+p)^2$, $s_L=(p_l+p_{\bar{\nu}})^2$,
$\theta$ ({\it i.e.\/}, the angle between the $D$ momentum in the $D\pi$ rest frame
and the moving direction of the $D\pi$ system in the $B$-meson's rest frame), 
$\theta_l$ ({\it i.e.\/}, the angle between the lepton momentum in the $l\bar{\nu}$
rest frame and the moving direction of the $l\bar{\nu}$ system in the
$B$-meson's rest frame) and $\phi$ ({\it i.e.\/}, the angle between the two decay planes
defined by the pairs (${\bf p},{\bf p}_D$) and 
(${\bf p}_l,{\bf p}_{\bar{\nu}}$) in the rest frame of the $B$-meson).

Since the initial $B^-$ system is not CP self-conjugate, any genuine
CP-odd observable can be constructed only by considering both the $B^-_{l4}$
decay and its charge-conjugated $B^+_{l4}$ decay
($B^+\to \overline{D^0} \pi^0 l^+ \nu_l$ or $B^+\to D^- \pi^+ l^+ \nu_l$), 
and by identifying the CP relations of their kinematic distributions.
The transition probability $|\overline{T}|^2$ for the $B^+$ decay in the same 
reference frame as in the $B^-$ decay is given by simple modification of the 
transition probability $|T|^2$ of the $B^-$ decay \cite{bcp}:
\be
|\overline{T}|^2=|T|^2\left\{\begin{array}{l}
   {\rm (i)\; change\; signs\; in\; front\; of\; the\; terms\; proportional\; 
    to\; imaginary\; part}\\
   {\rm (ii)\;}\zeta\to\zeta^*\;.\end{array}\right.
\label{ampbar}
\ee
We note that all the imaginary parts are being multiplied by
the quantity proportional to $\sin\phi$.
If the parameter $\zeta$ is real, the transition
probability for the $B^-$ decay and that of
the $B^+$ decay satisfy the following CP relation:
\be
|T|^2(\theta,\; \theta_l,\; \phi)=|\overline{T}|^2(\theta,\; \theta_l,\; -\phi).
\label{cprel}
\ee 
Then with CP violating complex parameter $\zeta$, 
$d\Gamma/d\Phi_4$ can be decomposed into a CP-even part ${\cal S}$ and
a CP-odd part ${\cal D}$:
\be
\frac{d\Gamma}{d\Phi_4}=\frac{1}{2}({\cal S}+{\cal D}).
\ee
The CP-even part ${\cal S}$ and the CP-odd part ${\cal D}$ can be easily
identified by making use of the CP relation (\ref{cprel}) between the $B^-$ and 
$B^+$ decay probabilities and they are expressed as
\be
{\cal S} =\frac{d(\Gamma+\overline{\Gamma})}{d\Phi_4},\;
{\cal D} =\frac{d(\Gamma-\overline{\Gamma})}{d\Phi_4},
\ee
where we have used the same kinematic variables $\{s_M,s_L,\theta,\theta_l\}$
for the $d\overline{\Gamma}/d\Phi_4$ except for the replacement of ${\phi}$ 
by $-\phi$, as shown in Eq. (\ref{cprel}).
Here $\Gamma$ and $\overline{\Gamma}$ are the decay rates for $B^-$ and $B^+$, 
respectively. The CP-even ${\cal S}$ term and the CP-odd ${\cal D}$ can be obtained from
$B^\mp$ decay probabilities and their explicit form is also listed in Ref.~\cite{bcp}.
We notice that the CP-odd term is proportional to the imaginary part 
of the parameter $\zeta$ in Eq.~(\ref{zeta}) and lepton mass $m_l$.
Therefore, there exists no CP violation in \bl decays within the SM 
since the SM corresponds to the case with $\zeta =0$. 

\section{Models}

\noindent
As possible new sources of CP violation detectable in the $B_{l 4}$ decays,
we consider new scalar-fermion interactions which preserve the 
symmetries of the SM. Then it can be proven that only four types of 
scalar-exchange models \cite{scalar} contribute to the 
$B\rightarrow D\pi l\bar{\nu}_l$.
One of them is the mult-Higgs-doublet (MHD) model \cite{Grossman} and
the other three models are scalar-leptoquark (SLQ) models \cite{wyler,randall}.
The authors of Ref.~\cite{tau} investigated CP violations in $\tau$ decay processes
within these extended models. We follow their description on the models and 
make it to be appropriate for our analysis.

\subsection{Multi-Higgs Doublet model}

\noindent
We first consider MHD model 
with $n$ Higgs doublets. The Yukawa 
interaction of the MHD model is
\be
{\cal L}_{MHD}=\bar{Q}_{L_i}F^D_{ij}\Phi_dD_{R_j}
              +\bar{Q}_{L_i}F^U_{ij}\tilde{\Phi}_uU_{R_j}
              +\bar{L}_{L_i}F^E_{ij}\Phi_eE_{R_j} + h.c.\;.
\ee
Here $Q_{L_i}$ denotes left-handed quark doublets, and $L_{L_i}$ denotes left-handed
lepton doublets, $D_{R_i}$ ($U_{R_i}$) and $E_{R_i}$ are for right-handed down (up)
quark singlets and right-handed charged lepton singlets, respectively.
The index $i$ is a generation index ($i=1,2,3$),
$\Phi_j$ ($j=1$ to $n$) are $n$ Higgs doublets and 
$\tilde{\Phi}_j=i\sigma_2\Phi^*_j$.
And among the $n$ doublets we denote by indices $d$, $u$ and $e$ 
the Higgs doublets that couple to down-type quarks, up-type quarks, 
and charged leptons, respectively.
$F^U$ and $F^D$ are general $3\times 3$ Yukawa matrices of which one matrix 
can be taken to be real and diagonal. Since neutrinos are massless, $F^E$ can be
chosen real and diagonal. The MHD model has $2(n-1)$ charged and $(2n-1)$ neutral
physical scalars. The Yukawa interactions of the $2(n-1)$ physical charged scalar
with fermions in the mass eigenstates read as
\be
{\cal L}_{MHD}=\sqrt{2\sqrt{2}G_F} \sum^n_{i=2}\Big[X_i(\bar{U}_L V {\cal M}_D D_R)+
     Y_i(\bar{U}_R {\cal M}_U V D_L)+ Z_i(\bar{N}_L{\cal M}_E E_R)\Big]H_i^+ +h.c.\;.
\ee
Here ${\cal M}_D$, ${\cal M}_U$ and ${\cal M}_E$ denote 
diagonal mass matrices of down-type quarks,
up-type quarks, and charged leptons, respectively.
$H_i^+$ are positively charged Higgs particles, $N_L$ are for left-handed neutrino
fields, and $V$ for the CKM matrix. $X_i$, $Y_i$ and $Z_i$ are complex coupling
constants which arise from the mixing matrix for charged scalars.

Within the framework of the MHD model, CP violation in charged scalar exchange can
arise for more than two Higgs doublets \cite{Wein,Glas}.
There are two mechanisms which give rise to CP violation in the scalar sector.
In one mechanism \cite{TDLee,Wein}, CP symmetry is maintained at the Lagrangian level
but broken through complex vacuum expectation values. However, this possibility has
been shown to have some phenomenological difficulties \cite{Pok,Grossman}.
In the other mechanism CP is broken by complex Yukawa couplings and possibly by 
complex vacuum expectation values so that CP violation can arise both from charged 
scalar exchange and from $W^\pm$ exchange. CP violation in both mechanisms is 
commonly manifest in phases that appear in the combinations $XY^*$, $XZ^*$ and $YZ^*$.

One crucial condition for CP violation in the MHD model is that not all the charged 
scalars should be degenerate. Then, without loss of generality and for simplicity,
we can assume that all but the lightest of the charged scalars effectively decouple 
from fermions. The couplings of the lightest charged scalar to fermions are
described by a simple Lagrangian
\be
{\cal L}_{MHD}=\sqrt{2\sqrt{2}G_F} \Big[X(\bar{U}_L V {\cal M}_D D_R)+
        Y(\bar{U}_R {\cal M}_U V D_L)+ Z(\bar{N}_L{\cal M}_E E_R)\Big]H^+ +h.c.\;.
\ee
This Lagrangian gives the effective Lagrangian contributing to the decay
$B\to D\pi l\bar{\nu}_l$,
\be
{\cal L}_{eff}=2\sqrt{2}G_FV_{cb}\frac{m_l}{M_H^2}\Big[m_b XZ^*(\bar{c}_L b_R)
            +m_c YZ^*(\bar{c}_R b_L)\Big](\bar{l}_R \nu_L),
\ee
at energies considerably low compared to the mass of the charged Higgs boson.
Then, one can show that the contribution from the MHD model to the \bl decay rate
is represented by the parameters:
\be                                            
\chi_{MHD}=0,\qquad \eta_{MHD}=\frac{m_l m_b}{M_H^2}\Big\{XZ^*-(\frac{m_c}{m_b})YZ^*\Big\},
\ee
and CP violation in the MHD model is determined by the parameter
\be
\Im(\zeta_{MHD})=\frac{m_l m_b}{M_H^2}\Big\{\Im(XZ^*)-(\frac{m_c}{m_b})\Im(YZ^*)\Big\}.
\label{xiMHD}
\ee
The constraint on the CP-violation parameter (\ref{xiMHD}) depend upon 
the values chosen for the $c$ and $b$ quark masses.
In the present work, we use
for the heavy $c$ and $b$ quark masses \cite{Godfrey},
\be
m_c=1628\;{\rm MeV},\qquad m_b=4977\;{\rm MeV}.
\ee
Clearly, sizable CP-violating effects require that $\Im(XZ^*)$ and
$\Im(YZ^*)$ are large and $M_H$ is small.
In the MHD model the strongest constraint \cite{Grossman} on $\Im(XZ^*)$
comes from the measurement of the branching ratio ${\cal B}(b\to X\tau
\nu_\tau)$ which actually gives a constraint on $|XZ|$. 
For $M_H<440$ GeV, the bound on $\Im(XZ^*)$ is given by
\be
\Im(XZ^*)\;<\;|XZ|\;<0.23M^2_H \;{\rm GeV}^{-2}.
\ee
On the other hand, the bound on $\Im(YZ^*)$ is mainly
given by $K^+\to \pi^+\nu\bar{\nu}$. The present bound \cite{Grossman} is
\be
\Im(YZ^*)\;<\;|YZ|\;<110.
\ee
Combining the above bounds, we obtain
the following bounds on $\Im(\zeta_{MHD})$ as
\be
&&|\Im(\zeta_{MHD})|\;<\; 1.88\;\qquad\qquad {\rm for}\;\tau\;{\rm family},\nonumber\\
&&|\Im(\zeta_{MHD})|\;<\; 0.11\;\qquad\qquad {\rm for}\;\mu\;{\rm family}.
\label{Hmodel}
\ee

\subsection{Scalar Leptoquark models}

\noindent
There are three types of SLQ models \cite{scalar,wyler} 
which can contribute to the $B\rightarrow D\pi l\nu$
at the tree level. The quantum numbers of the three leptoquarks under
the gauge group $SU(3)_C\times SU(2)_L\times U(1)_Y$ are
\be
&&\Phi_1=\left(3,2,\;\frac{7}{6}\right)\qquad {\rm (model\;I)},\nonumber\\
&&\Phi_2=\left(3,1,-\frac{1}{3}\right)\qquad {\rm (model\;II)},\nonumber\\
&&\Phi_3=\left(3,3,-\frac{1}{3}\right)\qquad {\rm (model\;III)},
\ee
respectively. The hypercharge $Y$ is defined to be $Q=I_3+Y$. The Yukawa 
couplings of the leptoquarks to fermions are given by
\be
&&{\cal L}^I_{SLQ}=[-x_{ij}\bar{Q}_{L_i}i\sigma_2E_{R_j}
                  +x'_{ij}\bar{U}_{R_i}L_{L_j}]\Phi_1+h.c.,\nonumber\\
&&{\cal L}^{II}_{SLQ}=[y_{ij}\bar{Q}_{L_i}i\sigma_2L^c_{L_j}
                  +y'_{ij}\bar{U}_{R_i}E^c_{R_j}]\Phi_2+h.c.,\nonumber\\
&&{\cal L}^{III}_{SLQ}=z_{ij}[\bar{Q}_{L_i}i\sigma_2\vec{\sigma}L^c_{L_j}]
                   \cdot\vec{\Phi}_3+h.c. .
\ee
Here the coupling constants $x^{(\prime)}_{ij},y^{(\prime)}_{ij}$, and $z_{ij}$
are complex when CP violation arises from the Yukawa interactions.
$\bar{Q}_{L_i}=(\bar{u}_i,\bar{d}_i)_L$ and $L_{L_i}=(\bar{\nu}_i,\bar{e}_i)_L$.
The superscript $c$ denotes charge conjugation, i.e., 
$\psi^c_{R,L}=i\gamma^0\gamma^2\bar{\psi}^T_{R,L}$ for a spinor field $\psi$.
$\vec{\sigma}=(\sigma_1,\sigma_2,\sigma_3)$ and $\sigma_i(i=1,2,3)$ are
the Pauli matrices. In terms of the charge component of the leptoquarks,
the Lagrangian relevant to the 
$B\rightarrow D\pi l\bar{\nu}_l$ decay is given by
\be
&&{\cal L}^I_{SLQ}=[x_{3j}\bar{b}_Ll_R+x'_{2j}\bar{c}_R\nu_{lL}]
                    \phi_1^{(2/3)} + h.c.,\nonumber\\
&&{\cal L}^{II}_{SLQ}=[-y_{3j}(\bar{t}_Ll^c_L-\bar{b}_L\nu^c_{lL})
                 +y'_{2j}\bar{c}_R l^c_R]\phi_2^{(-1/3)} + h.c.,\nonumber\\
&&{\cal L}^{III}_{SLQ}=-[z_{2j}(\bar{c}_Ll^c_L+\bar{s}_L\nu^c_{lL})
    +z_{3j}(\bar{t}_L l^c_L+\bar{b}_L\nu^c_{lL})]\phi_3^{(-1/3)} + h.c.\;,
\ee
where $j=2,3$ for $l=\mu,\;\tau$, respectively.
After Fierz rearrangement we obtain the effective SLQ Lagrangians which can give 
contribution to $B_{l4}$ decay:
\be
{\cal L}^I_{eff}&=&-\frac{x_{3j}x^{\prime *}_{2j}}{2M^2_{\phi_1}}\left[
             (\bar{b}_Lc_R)(\bar{\nu}_{lL}l_R)
            +\frac{1}{4}(\bar{b}_L\sigma^{\mu\nu}c_R)
             (\bar{\nu}_{lL}\sigma_{\mu\nu}l_R)\right]+h.c.,\nonumber\\
{\cal L}^{II}_{eff}&=&-\frac{y_{3j}y^{\prime *}_{2j}}{2M^2_{\phi_2}}\left[
             (\bar{b}_Lc_R)(\bar{l}^c_R\nu^c_{l L})
            +\frac{1}{4}(\bar{b}_L\sigma^{\mu\nu}c_R)
          (\bar{l}^c_R\sigma_{\mu\nu}\nu^c_{l L})\right]\nonumber\\
          &&+\frac{y_{3j}y^*_{2j}}{2M^2_{\phi_2}}(\bar{b}_L\gamma_\mu c_L)
           (\bar{l}^c_L\gamma^\mu\nu^c_{l L})+h.c.,\nonumber\\
{\cal L}^{III}_{eff}&=&-\frac{z_{3j}z^*_{2j}}{2M^2_{\phi_3}}(\bar{b}_L\gamma_\mu c_L)
           (\bar{l}^c_L\gamma^\mu\nu^c_{l L})+h.c.
\ee
The tensor parts in Model I and Model II may contribute to the 
$B\rightarrow D\pi l\bar{\nu}_l$ decay through the spin-2 resonances
such as $D^*_2$. In the present work, however, spin-2 resonances do not enter, 
so we will neglect these
tensor contributions, which take the same form 
and have the same couplings as the scalar resonance contributions.

Then the size of new contributions from these three SLQ models is parametrized as
\be
&&\chi^I_{SLQ}=0,\qquad\qquad\qquad\qquad \eta^I_{SLQ}=
     -\frac{x^*_{3j}x'_{2j}}{4\sqrt{2}G_FV_{cb}M^2_{\phi_1}},\nonumber\\
&&\chi^{II}_{SLQ}=-\frac{y_{3j}y^*_{2j}}{4\sqrt{2}G_FV_{cb}M^2_{\phi_2}},\quad\; 
 \eta^{II}_{SLQ}=-\frac{y^*_{3j}y'_{2j}}{4\sqrt{2}G_FV_{cb}M^2_{\phi_2}},\nonumber\\
&&\chi^{III}_{SLQ}=\frac{z_{3j}z^*_{2j}}{4\sqrt{2}G_FV_{cb}M^2_{\phi_3}},\qquad  
 \eta^{III}_{SLQ}=0.
\ee
Note that Model I acquires the contribution only from the scalar-type interaction.
On the other hand, Model III dose only from the vector-type interactions, so
the Model III does not contribute to CP violation in the $B_{l4}$ decay.
The terms with $y_{3j}y^*_{2j}$ in Model II and $z_{3j}z^*_{2j}$ 
in Model III modify the size of the vector contributions.
Taking the approximation that these new contributions are negligible
compared to the SM contributions, we find that the size of the SLQ models
CP-violation effects is dictated by the CP-odd parameters
\begin{eqnarray}
&&{\rm Im}(\zeta^I_{\rm SLQ})\approx-\frac{{\rm Im}[x_{3j}
      x^{\prime *}_{2j}]}{4\sqrt{2}G_FV_{cb}M^2_{\phi_1}}\,,\nonumber\\
&&{\rm Im}(\zeta^{II}_{\rm SLQ})\approx -\frac{{\rm Im}[y_{3j}y^{\prime *}_{2j}]}
   {4\sqrt{2}G_FV_{cb}M^2_{\phi_2}}\,,\nonumber\\
&&{\rm Im}(\zeta^{III}_{\rm SLQ})\approx 0\,.
\label{LQp}
\end{eqnarray}
This approximation is justified because the contributions from new physics 
are expected to be very small compared to those from the SM. 

 Although there are at present no direct 
constraints on the SLQ models CP-odd parameters in (\ref{LQp}), 
a rough constraint to the parameters can be provided on the assumption 
\cite{Davidson} that $|x^\prime_{2j}|\sim |x_{2j}|$ and 
$|y^\prime_{2j}|\sim |y_{2j}|$, that is to say, the leptoquark couplings   
to quarks and leptons belonging to the same generation are of a similar
size; then the experimental upper bound 
from $B\rightarrow l\bar{l}X$ decay for Model I, and $B\rightarrow l\nu X$
for Model II yields \cite{Davidson}
\begin{eqnarray}
&&|{\rm Im}(\zeta^I_{\rm SLQ})|< 0.38,\quad 
|{\rm Im}(\zeta^{II}_{\rm SLQ})|< 1.52 \qquad\; {\rm for}\;\tau\;{\rm family},\nonumber\\
&&|{\rm Im}(\zeta^I_{\rm SLQ})|< 0.03,\quad 
|{\rm Im}(\zeta^{II}_{\rm SLQ})|< 1.52 \qquad\; {\rm for}\;\mu\;{\rm family}.
\label{Models}
\end{eqnarray}
Based on the constraints (\ref{Models}) to the CP-odd
parameters, we quantitatively estimate the number of the semileptonic
$B_{l4}$ decays to detect CP violation for the maximally-allowed values 
of the CP-odd parameters.

\section{Asymmetries}

\noindent
An easily-constructed CP-odd asymmetry is the rate asymmetry
\be
A\equiv \frac{\Gamma-\overline{\Gamma}}{\Gamma+\overline{\Gamma}},
\ee
where $\Gamma$ and $\overline{\Gamma}$ are the decay rates for $B^-$ and $B^+$, 
respectively.
The statistical significance of the asymmetry can then be computed as
\be
N_{SD}=\frac{N_- -N_+}{\sqrt{N_-+N_+}} =\frac{N_- -N_+}{\sqrt{N \cdot Br}},
\ee
where $N_{SD}$ is the number of standard deviations,
$N_\pm$ is the number of events predicted in $B_{l4}$ decay for $B^\pm$
meson, $N$ is the number of $B$ meson produced,
and $Br$ is the branching fraction of the relevant $B$ decay mode. 
Taking $N_{SD}=1$, we obtain 
the number $N_B$ of the B mesons needed to observe CP violation at $1$-$\sigma$ level:
\be
N_B=\frac{1}{Br\cdot A^2}.
\ee

Next, we consider the so-called optimal observable.
An appropriate real weight function $w(s_M,s_L;\theta,\theta_l,\phi)$
is usually employed to separate the CP-odd ${\cal D}$ contribution and to enhance
its analysis power for the CP-odd parameter ${\rm Im}(\zeta)$ through 
the CP-odd quantity:
\begin{eqnarray}
\langle w{\cal D}\rangle\equiv\int\left[w{\cal D}\right] d\Phi_4,
\end{eqnarray}
of which the analysis power is determined by the parameter 
\begin{eqnarray}
\varepsilon
   =\frac{\langle w{\cal D}\rangle}{\sqrt{\langle{\cal S}\rangle
          \langle w^2{\cal S}\rangle}}\;.
\label{Significance}
\end{eqnarray}
For the analysis power $\varepsilon$, the number $N_B$ of the $B$ mesons  
needed to observe CP violation at 1-$\sigma$ level is
\begin{eqnarray}
N_B=\frac{1}{Br\cdot\varepsilon^2}\;.
\label{eq:number}
\end{eqnarray}
Certainly, it is desirable to find the optimal weight function
with the largest analysis power. It is known  \cite{Optimal} that 
when the CP-odd contribution to the total rate is relatively small, 
the optimal weight function  is approximately given by
\begin{eqnarray}
w_{\rm opt}(s_M,s_L;\theta,\theta_l,\phi)=
 \frac{{\cal D}}{{\cal S}}~~~\Rightarrow~~~
 \varepsilon_{\rm opt}=\sqrt{\frac{\langle\frac{{\cal D}^2}{{\cal S}}\rangle}
 {\langle{\cal S}\rangle}}.
\end{eqnarray}
We adopt this optimal weight function in the following numerical analyses.

\section{Numerical Results and Conclusions}

\noindent
We first consider decay to heavy lepton, $\tau$, since CP-odd asymmetry is 
proportional to $m_l$, and heavy lepton may be more susceptible to effects
of new physics.
We restrict ourselves to the soft-pion limit by considering only the region
$s_M\le 6.5\;{\rm GeV}^2$ which is about one half of the maximum value.
This restriction corresponds to the pion momentum less than about $0.6$ GeV.

In Table 1, we show the results of $B_{\tau 4}$ decays 
for the two CP-violating asymmetries;
the rate asymmetry $A$ and the optimal asymmetry $\varepsilon_{\rm opt}$.
We estimated the number of $B$ meson pairs, $N_B$, needed for detection at 
$1\sigma$ level for maximally-allowed values of CP-odd parameters in
Eq.~(\ref{Hmodel}) and (\ref{Models}).
As we can expect, the optimal observable gives much better result than 
the simple rate asymmetry. We also found that although the decay rate 
of neutral pion
mode is larger than that of the charged pion mode 
because the most dominant resonance $D^{0*}$
can not decay into $D^+\pi^-$ in its mass-shell, 
the latter case gives better detection results
because of large CP-violating effects in charged pion decay mode.
Since one expect about $10^8$ orders of $B$ meson pairs produced yearly in the 
asymmetric $B$ factories, one could probe the CP-violation effects
down to the current bounds of the MHD model and SLQ models
by using the optimal asymmetry observable.

We also estimated CP-violation effects in the \bl decays with light leptons.
The results for $B_{\mu 4}$ decays are shown in Table 2.
We found that $B_{\mu 4}$ decay modes give worse results than $B_{\tau 4}$ cases.
In Ref.~\cite{bcp}, we found that both $B_{\tau 4}$ and $B_{\mu 4}$ decay modes
could be equally good probes for the same values of CP-odd parameter $\Im(\zeta)$.
At the same time, we discussed that in some extensions of 
the SM the CP-odd parameter itself would be proportional to lepton mass. 
And therefore, $B_{\tau 4}$ decay modes 
would serve as definitely much better probes of CP violation than light 
lepton cases in those models.
This is the case in the MHD model where
$\Im(\zeta_{MHD})=1.88\;(0.11)$ for $\tau\;(\mu)$ family, directly resulted from its
dependence on lepton mass.
But there's no such dependence in SLQ models.
Although the numerical values of CP-odd parameters in SLQ model I,
$\Im(\zeta^I_{SLQ})=0.38\;(0.03)$ for $\tau\;(\mu)$ family, look like there's
some lepton mass dependence in it,
actually different values are from current experimental bounds.
For instance, the constraints on the CP-odd parameter in SLQ model I come from
$B\to \tau\bar{\tau}X$ for $\tau$ family and $B\to \mu\bar{\mu}X$ for $\mu$ family,
respectively, and so the smaller CP-odd value for $\mu$ family implies
current experimental constraint on muon mode is more strict.
For the SLQ model II, however, current bounds are roughly same for $\tau$ and $\mu$ families.
Therefore, if one use optimal asymmetry observable,
$B_{\tau 4}$ decay modes would provide more stringent constraints to all the extended models
we have considered, and one could also use $B_{\mu 4}$ decays to constrain at least SLQ model II.

In conclusions, we investigated
CP violation from physics beyond the Standard Model through semileptonic $B_{l4}$
decays: $B\to D\pi l\bar{\nu}_l$. We considered as new sources of CP violation 
multi-Higgs doublet model and scalar-leptoquark models.
In the decay process, we included various excited states as intermediate states
decaying to the final hadrons. The CP violation is implemented 
through interference between $W$-exchange diagrams and 
Scalar-exchange diagrams with complex couplings in the extended models above.
We calculated the CP-odd rate asymmetry and the optimal asymmetry 
for $B_{\tau 4}$ and $B_{\mu 4}$ decay modes, and
found that the optimal asymmetry for $B_{\tau 4}$ decays is sizable and
can be detected at $1\sigma$ level with about $10^6$-$10^7$ orders of $B$-meson pairs, 
for maximally-allowed values of CP-odd parameters in the models of our interest.
Since $\sim$$10^8$ $B$-meson pairs are expected to be produced yearly 
at the forthcoming $B$ factories,
one could investigate CP-violation effects by using our formalism.
In other words, one could give more stringent constraints to all the models we have considered
using $B_{\tau 4}$ decay modes.
\\

\section*{Acknowledgments}

\noindent
CSK wishes to thank the Korea Institute for Advanced Study for warm
hospitality.
CSK wishes to acknowledge the financial support of 
Korean Research Foundation made in the program of 1997.
The work of JL was supported 
in part by Non-Directed-Research-Fund of 1997, KRF,
in part by the CTP, Seoul National University, 
in part by the BSRI Program, Ministry of Education, Project No. BSRI-98-2425,
in part by the KOSEF-DFG large collaboration project, 
Project No. 96-0702-01-01-2.
The work of WN was 
supported by the BSRI Program, 
Ministry of Education, Project No. BSRI-98-2425.

\vskip 2cm 
%

\begin{table}
{Table~1}. {The CP-violating rate asymmetry A and the optimal asymmetry
$\varepsilon_{\rm opt}$, determined in the soft pion limit,  
and the number of charged B meson pairs, 
$N_B$, needed for detection at $1\sigma$ level,
at reference values $\Im(\zeta_{MHD})=1.88$, $\Im(\zeta^I_{SLQ})=0.38$ and 
$\Im(\zeta^{II}_{SLQ})=1.52$,} for the $B_{\tau 4}$ decays.\par
\begin{tabular}{c|cc|cc|cc}
\multicolumn{7}{c}{(a) $B^-\to D^+\pi^-\tau\bar{\nu}_\tau$ mode}\\
\hline
Model
&\multicolumn{2}{c|}{MHD}&\multicolumn{2}{c|}{SLQ I}&\multicolumn{2}{c}{SLQ II}\\
\hline\hline 
Asymmetry & Size(\%) & $N_B$ & Size(\%) & $N_B$ & Size(\%) & $N_B$\\
\hline
A & $0.31$ & $2.34\times 10^{8}$ & $0.06$ & $5.74\times 10^{9}$ & $0.25$ & $3.59\times 10^{8}$\\
$\varepsilon_{\rm opt}$ & $4.25$ & $1.25\times 10^6$ & $0.86$ & $3.06\times 10^7$ & $3.44$ & $1.91\times 10^{6}$\\
\hline
\multicolumn{7}{c}{ }\\
\hline
\multicolumn{7}{c}{(b) $B^-\to D^0\pi^0\tau\bar{\nu}_\tau$ mode}\\
\hline
Model
&\multicolumn{2}{c|}{MHD}&\multicolumn{2}{c|}{SLQ I}&\multicolumn{2}{c}{SLQ II}\\
\hline\hline 
Asymmetry & Size(\%) & $N_B$ & Size(\%) & $N_B$ & Size(\%) & $N_B$\\
\hline
A & $0.006$ & $2.28\times 10^{10}$ & $0.001$ & $5.58\times 10^{11}$ & $0.005$ & $3.49\times 10^{10}$\\
$\varepsilon_{\rm opt}$ & $0.6$ & $2.36\times 10^6$ & $0.1$ & $5.77\times 10^7$ & $0.45$ & $3.61\times 10^{6}$\\
\end{tabular}
\end{table}
\begin{table}
{Table~2}. {The CP-violating rate asymmetry A and the optimal asymmetry
$\varepsilon_{\rm opt}$, determined in the soft pion limit, 
and the number of charged B meson pairs, $N_B$, needed for detection at $1\sigma$ level
at reference values $\Im(\zeta_{MHD})=0.11$, $\Im(\zeta^I_{SLQ})=0.03$ and 
$\Im(\zeta^{II}_{SLQ})=1.52$,} for the $B_{\mu 4}$ decays.\par
\begin{tabular}{c|cc|cc|cc}
\multicolumn{7}{c}{(a) $B^-\to D^+\pi^-\mu\bar{\nu}_\mu$ mode}\\
\hline
Model
&\multicolumn{2}{c|}{MHD}&\multicolumn{2}{c|}{SLQ I}&\multicolumn{2}{c}{SLQ II}\\
\hline\hline 
Asymmetry & Size(\%) & $N_B$ & Size(\%) & $N_B$ & Size(\%) & $N_B$\\
\hline
A & $0.003$ & $1.5\times 10^{11}$ & $0.0008$ & $2.02\times 10^{12}$ & $0.04$ & $7.87\times 10^{8}$\\
$\varepsilon_{\rm opt}$ & $0.05$ & $4.87\times 10^8$ & $0.014$ & $6.54\times 10^9$ & $0.72$ & $2.55\times 10^{6}$\\
\hline
\multicolumn{7}{c}{ }\\
\hline
\multicolumn{7}{c}{(b) $B^-\to D^0\pi^0\mu\bar{\nu}_\mu$ mode}\\
\hline
Model
&\multicolumn{2}{c|}{MHD}&\multicolumn{2}{c|}{SLQ I}&\multicolumn{2}{c}{SLQ II}\\
\hline\hline 
Asymmetry & Size(\%) & $N_B$ & Size(\%) & $N_B$ & Size(\%) & $N_B$\\
\hline
A & $0.0001$ & $6.07\times 10^{12}$ & $0.00004$ & $8.15\times 10^{13}$ & $0.002$ & $3.18\times 10^{10}$\\
$\varepsilon_{\rm opt}$ & $0.01$ & $9.37\times 10^8$ & $0.003$ & $1.26\times 10^{10}$ & $0.15$ & $4.91\times 10^{6}$\\
\end{tabular}
\end{table}
%
\newpage
\parindent=0 cm
\begin{figure}[h]
\vspace*{-2.0cm}
\hbox to\textwidth{\hss\epsfig{file=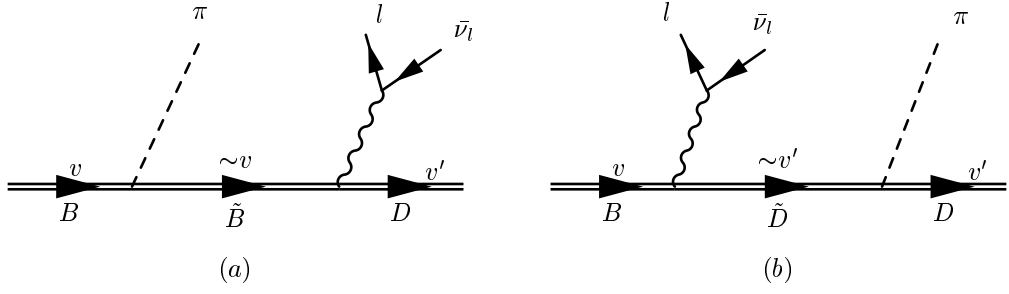,height=30cm}\hss}
\vspace*{-21.5cm}
\caption{Feynman diagrams for $B_{l4}$ decays.}
\label{fig:diagram}
\end{figure}

\end{document}